\documentclass[%
 reprint,
 amsmath,amssymb,
 aps,
prd,
floatfix,
]{revtex4-2}

\usepackage{graphicx}
\usepackage{dcolumn}
\usepackage{bm}
\usepackage{slashed}
\usepackage{color}
\usepackage{textcomp}
\usepackage{enumerate}
\usepackage{comment}
\usepackage{float}

\usepackage{tikz}
\usetikzlibrary{calc} 
\usetikzlibrary{patterns} 
\usetikzlibrary{decorations.pathreplacing} 
\usetikzlibrary{decorations.markings} 
\usetikzlibrary{decorations.pathmorphing} 
\usetikzlibrary{positioning}

\newcommand{\aref}[1]{App.~\ref{#1}}
\newcommand{\eq}[1]{Eq.~\eqref{#1}}

\newcommand{\eqs}[2]{Eqs.~\eqref{#1} and \eqref{#2}}
\newcommand{\eqt}[2]{Eqs.~(\ref{#1}-\ref{#2})}
\newcommand{\fref}[1]{Fig.~\ref{#1}}

\newcommand{\sref}[1]{Sec.~\ref{#1}}

\newcommand{\dd}{\mathrm{d}}

\newcommand{\bs}{\boldsymbol}
\newcommand{\nn}{\nonumber}

\def\mnras{Mon. Not. R. Astron. Soc.}

\def\apj{Astrophys. J.}
\def\apjl{Astrophys. J. Lett.}
\def\apjs{Astrophys. J., Suppl. Ser.}
\def\prd{Phys. Rev. D}
\def\prl{Phys. Rev. Lett.}

\def\aap{Astron. Astrophys.}

\def\araa{Annu. Rev. Astron. Astrophys.}

\def\baas{Bull. Am. Astron. Soc.}
\def\nat{Nature}

\begin{document}

\preprint{APS/123-QED}

\title{Moving gravitational wave sources at cosmological distances: Impact on the measurement of the Hubble constant}

\author{Alejandro Torres-Orjuela}
\affiliation{MOE Key Laboratory of TianQin Mission, TianQin Research Center for Gravitational Physics \& School of Physics and Astronomy, Frontiers Science Center for TianQin, Gravitational Wave Research Center of CNSA, Sun Yat-Sen University (Zhuhai Campus), Zhuhai 519082, China}

\author{Xian Chen}
\email{Corresponding author: xian.chen@pku.edu.cn}
\affiliation{Astronomy Department, School of Physics, Peking University, 100871 Beijing, China}
\affiliation{Kavli Institute for Astronomy and Astrophysics at Peking University, 100871 Beijing, China}

\date{\today}

\begin{abstract}
Standard sirens -- gravitational wave (GW) sources with an electromagnetic (EM) counterpart -- can be used to measure the Hubble constant directly which should help to ease the existing Hubble tension. However, if the source has a relative velocity to the expanding universe on top of its motion due to the Hubble flow, a relativistic redshift affects the redshift of the EM counterpart and the apparent distance of the GW source, and thus it needs to be corrected to obtain accurate measurements. We study the effect of such a relative velocity on GWs for a source in an expanding universe showing that the total redshift of the wave is equal to the product of the relativistic redshift and the cosmological redshift. We, further, find that a relative velocity of the source changes its apparent distance by a factor $(1+z_{\rm rel})^2$ in contrast to a linear factor for the cosmological redshift. We discuss that the additional factor for the relativistic redshift is a consequence of a velocity-dependent amplitude for GWs. We consider the effect of the relative velocity on the chirp mass and the apparent distance of the source an observer would infer when ignoring this velocity. We find that for different astrophysical scenarios the error, i.e., the deviation between the value inferred and the actual value, can range between 0.1\,\% and 7\,\% for the chirp mass while the error in the apparent distance can be between 0.25\,\% and 15\,\%. Furthermore, we consider the error introduced in the measurement of the Hubble constant using standard sirens for two cases: (i) when the effect of velocity on the redshift of the EM counterpart is considered but not on the apparent distance obtained from GWs and (ii) when the effect of the relative velocity is ignored completely. We find that in the first case the error can reach 1\,\% for a source moving due to the peculiar velocity of its host galaxy and that in the second case the error can be more than 5\,\% for a source at the distance of GW150914 with the same velocity.
\end{abstract}


\maketitle

\section{Introduction}\label{sec:int}

With almost 100 gravitational wave (GW) detections up to date, we can firmly say that the era of GW astronomy has started~\cite{GWTC1,GWTC2,GWTC3} and it will only become more intriguing with the improving accuracy of current a future gravitational wave detectors~\cite{ligo_2015,virgo_2012,kagra_2019,lisa_2017,tianqin_2016,taiji_2015,et_2010,cosmic_explorer_2019,tse_yu_2019}. A particularly interesting application in GW astronomy is to use standard sirens -- GW events with an electromagnetic (EM) counterpart -- to measure the expansion of the universe~\cite{schutz_1986,holz_hughes_2005,ligo_2017d,ligo_2017c,chen_fishbach_2018,mukherjee_ghosh_2020,ligo_2021d,chen_haster_2022}. For such a source, we can use the luminosity distance from GW detection together with the redshift from its EM counterpart to determine the Hubble constant. However, the relation between the Hubble constant, a redshift of the source, and its luminosity distance is derived under the assumption that the source is at rest relative to the expanding universe and thus a relative velocity of the source on top of the Hubble flow needs to be corrected when inferring the Hubble constant~\cite{dodelson_2020}.

Different astrophysical scenarios suggest that GW sources are moving with relative velocities ranging from a few $100\,{\rm km\,s^{-1}}$ up to several percent of the speed of light~\cite{lisa_2022}. Three particularly interesting cases that affect most if not all GW sources are (i) gravitational kicks induced by the merger of the source~\cite{pretorius_2005,herrmann_hinder_2007a,herrmann_hinder_2007b,herrmann_hinder_2007c,koppitz_pollney_2007,baker_centrella_2006,baker_boggs_2008,campanelli_lousto_2006,campanelli_lousto_2007a,campanelli_lousto_2007b,healy_herrmann_2009}, (ii) binary black holes (BBHs), binary neutron stars (BNSs), and black hole neutron star binaries (BHNSB) orbiting a supermassive black hole (SMBH) in the center of a galaxy~\cite{antonini_perets_2012,mckernan_ford_2012,bartos_kocsis_2017,stone_metzger_2017,chen_han_2018,chen_li_2019,addison_gracia-linares_2019,tagawa_haiman_2020,pan_lyu_2021,derdzinski_mayer_2022}, and (iii) a peculiar velocity of the host galaxy~\cite{zinn_west_1984,bahcall_1988,girardi_fadda_1996,carlberg_yee_1996,springel_white_2001,ruel_bazin_2014,scrimgeour_davis_2016,colin_mohayaee_2017}. In particular, the peculiar velocity of the host galaxy has been considered in the literature when correcting for the effect of velocity~\cite{wang_wang_2018,howlett_davis_2020,nicolaou_lahav_2020,mukherjee_lavaux_2021,ligo_2021d}. However, the corrections only focus on the effect the relative velocity has on the redshift of the EM counterpart but ignore the effect of this velocity on the apparent distance.

Different methods have been proposed to measure the relative velocity of a GW source by either considering a time-dependent phase shift~\cite{gerosa_moore_2016,meiron_kocsis_2017,inayoshi_tamanini_2017,tamanini_klein_2019,wong_baibhav_2019,torres-orjuela_chen_2020,yu_chen_2021,strokov_fragione_2022} or a change in the spherical modes of the GW~\cite{calderon-bustillo_clark_2018,varma_gerosa_2019,varma_biscoveanu_2020,varma_isi_2020,torres-orjuela_chen_2021,torres-orjuela_amaro-seoane_2021}. From the information of the source's relative velocity from GWs, we can correct the apparent distance for its change due to the relativistic redshift as well as set tighter constraints on the redshift of the EM counterpart. Taking these two corrections into account will greatly reduce the estimation error, i.e., the difference between the value inferred from detection and the actual value, for the Hubble constant thus helping to ease the Hubble tension~\cite{di-valentino_mena_2021,dainotti_de-simone_2021,perivolaropoulos_skara_2021,jedamzik_pogosian_2021,vagnozzi_2021,krishnan_mohayaee_2021,freedman_2021,di-valentino_anchordoqui_2021,rameez_sarkar_2021,dainotti_de-simnoe_2022,vagnozzi_pacucci_2022,mortsell_goobar_2022}. Therefore, it is crucial to have a detailed understanding of the effect of a source's relative velocity on the GW's frequency and the obtained apparent distance in an expanding universe.

In \sref{sec:crs}, we start revising the effect of an expansion of the universe on the frequency of GWs and the luminosity distance of the source. We then study the effect of a source's relative velocity on the frequency and the apparent distance of the source in a static universe in \sref{sec:rrs} to then extend our analysis in \sref{sec:trs} to the case of a source moving with a relative velocity on top of the Hubble flow in an expanding universe. Most remarkable, we find that the apparent distance for a source with a relative velocity scales with the square of the relativistic redshift, in contrast to the case of cosmological redshift where the distance scales with a linear factor of the redshift. In \sref{sec:obs}, we analyze the implication of our results on the estimated chirp mass and distance of the source, where we show that for realistic astrophysical scenarios the estimation error of the two can go up to several percent. We, further, discuss the square factor of the relativistic redshift in the apparent distance and show that it can be attributed to a velocity-dependent change in the amplitude of GWs. We study the implications of our results for the measurement of the Hubble constant using standard sirens for the case where only the redshift of the EM counterpart is considered and when the relative velocity of the source is completely ignored. We find that for a typical peculiar velocity of the source the error can already reach 1\,\% for the first case while for the second case the error can go up to several 10\,\% depending on the distance of the source. We finalise drawing conclusions in \sref{sec:con}. Throughout this paper, unless otherwise indicated, we use geometrical units in which the gravitational constant and the speed of light are equal to one (i.e., $G=c=1$). Furthermore, we use bold symbols for three-dimensional spatial vectors and a tilde to mark four-dimensional space-time vectors.

\section{Cosmological redshift}\label{sec:crs}

We start considering the effect of the cosmological expansion on the frequency of a GW in a Friedmann-Lemaître-Robertson-Walker (FLRW) universe, described by the metric~\cite{misner_thorne_1973}
\begin{equation}\label{eq:flrw}
    \dd s^2 = \dd t^2 - a^2(t)\left(\frac{\dd r^2}{1-kr^2} + r^2\dd\theta^2 + r^2\sin^2(\theta)\dd\phi^2\right).
\end{equation}
Here, $a(t)$ is the time-dependent scale factor that describes the expansion of the universe while $k$ describes its curvature, where $k=0$ corresponds to a flat universe, $k>0$ corresponds to a closed universe, and $k<0$ corresponds to an open universe. For simplicity and because observational data suggests that our universe is relatively flat~\cite{planck_2020,spergel_bean_2007}, we restrict the analysis in this paper to the case of a flat universe $k=0$. However, in principle, the methods developed in this paper can be extended to the cases of a close and an open universe with some additional effort. For the redshift of a closed or an open universe when only considering the cosmological expansion see, e.g., Ref.~\cite{maggiore_2008}.

For the derivation of the cosmological redshift of GWs, we follow the procedure introduced in Ref.~\cite{maggiore_2008} where we consider a source located at a comoving distance $r_s$ and an observer located at $r=0$. Suppose the source emits a wave with a crest at a time $t_s$ and a second crest at a time $t_s+\Delta t_s$. Further, assume the first cress is detected by the observer at a time $t_o$ while the second crest is detected at a time $t_o+\Delta t_o$. Because in general relativity GWs travel at the speed of light~\cite{misner_thorne_1973}, we can describe the path of the first crest by setting $\dd s^2=0$ in \eq{eq:flrw} and solving the integrals
\begin{equation}\label{eq:cri1}
    \int_{t_s}^{t_o}\frac{\dd t}{a(t)} = -\int_0^{r_s}\dd r,
\end{equation}
while for the second crest, we have to solve the integrals
\begin{equation}\label{eq:cri2}
    \int_{t_s+\Delta t_s}^{t_o+\Delta t_o}\frac{\dd t}{a(t)} = -\int_0^{r_s}\dd r.
\end{equation}
Note, that in both cases the integral over $r$ is the same because in comoving coordinates the coordinate locations of the source and the observer do not change if they are initially at rest~\cite{maggiore_2008}.

Solving the integrals, then subtracting the results of \eq{eq:cri1} from the results of \eq{eq:cri2}, and expanding to linear order in $\Delta t_s$ and $\Delta t_o$, we find
\begin{equation}\label{eq:ctd1}
    \Delta t_o = \frac{a(t_o)}{a(t_s)}\Delta t_s.
\end{equation}
Therefore, we find that in an expanding universe there is a dilation of the time measured by an observer far from the source. Using the usual convention that ${a(t_o)=1}$~\cite{dodelson_2020}, we define the cosmological redshift as
\begin{equation}\label{eq:dcrs}
    (1+z_{\rm cos}) := a^{-1}(t_s),
\end{equation}
leading to the following relation for the time measured by the observer's and source's clocks
\begin{equation}\label{eq:ctd2}
    \dd t_o = (1+z_{\rm cos})\dd t_s.
\end{equation}
Using this result and that the frequency of a wave is the time derivative of the coordinate invariant phase of the wave, we get that the frequency in the observer frame $f_o$ and the frequency in the source frame $f_s$ fulfill
\begin{equation}\label{eq:crs}
    f_o = \frac{f_s}{(1+z_{\rm cos})}.
\end{equation}

After deriving the effect of cosmological expansion on the frequency of a GW, we analyze how the apparent distance of the source $D_{\rm app}$ is affected by the cosmological redshift. The apparent distance can be determined using the energy flux $F$ of the source that corresponds to the energy $E$ irradiated per unit time $\dd t$ per unit area $\dd A$
\begin{equation}\label{eq:flux}
    F := \frac{\dd E}{\dd A\dd t}.
\end{equation}
The apparent distance of a GW is then defined as the radius of a sphere for which the absolute luminosity of the source $L := \dd E_s/\dd t_s$ equals the flux in a unit area
\begin{equation}\label{eq:dad}
    F = \frac{L}{D_{\rm app}^2\dd\Omega_s},
\end{equation}
where we used that for a sphere $\dd A = r^2\dd\Omega$. In the case of cosmological expansion, the apparent distance is equal to the luminosity distance $D_L$~\cite{maggiore_2008} but we use the term apparent distance to distinguish the two in the more general case where the source is not only affected by cosmological expansion.

To get the apparent distance for a GW source in an expanding universe, we use that the energy of a GW transforms in the same way as its frequency~\cite{maggiore_2008}
\begin{equation}\label{eq:ecr}
    E_o = \frac{E_s}{(1+z_{\rm cos})},
\end{equation}
that the time is dilated as in \eq{eq:ctd2}, and that $\dd\Omega$ is not affected by the cosmological expansion because it is isotropic while the distance between the source and the observer is $a(t_o)r_d$. However, using that $a(t_0)=1$, we have that the distance between the source and the observer is equal to the comoving distance $r_d$. Putting everything together, we obtain
\begin{equation}\label{eq:fadc}
    F = \frac{L}{(1+z_{\rm cos})^2r_d^2\dd\Omega_s}
\end{equation}
and thus find the apparent distance
\begin{equation}\label{eq:adc}
    D_{\rm app} = (1+z_{\rm cos})r_d.
\end{equation}
Therefore, we recover the classical result that the apparent distance, which in this case is equal to the luminosity distance $D_L$, equals the comoving distance times the same factor by which the frequency is redshifted~\cite{sathyaprakash_schutz_2009}.

\section{Relativistic redshift}\label{sec:rrs}

In this section, we derive the relativistic redshift and the change of the apparent distance induced by a relative velocity between the source and the observer in a static universe using the same methods as in the previous section. Therefore, we consider a Minkowsky space with the metric~\cite{schutz_2009}
\begin{equation}\label{eq:mink}
    \dd s^2 = \dd t^2 - \dd r^2 - r^2\dd\theta^2 - r^2\sin^2(\theta)\dd\phi^2.
\end{equation}
The case of a moving source in a flat but expanding FLRW universe is discussed in the next section.

Before we derive the effect of the relative velocity, we need to fix the coordinates systems (COs) for the observer $O$ and the source $S$. We set the origins of the two COs to agree at the time $t=t'=0$, where we mark quantities in $S$ with a prime while quantities in $O$ remain unprimed. In $O$, the source moves with a relative velocity $\bs{v}$ while the observer remains at the origin of the CO and in $S$ the observer moves with a relative velocity $\bs{v}'$ while the source remains at the radial distance $r'_s$. To simplify the calculations, we set $O$ to have its $z$-axis parallel to the velocity of the source $\bs{v}$ and $S$ to have its $z$-axis anti-parallel to the velocity of the observer $\bs{v}'$. With this setting the two COs are parallel and the polar angle $\theta$ is affected by the relative velocity while the azimuthal angle $\phi$ is not.

We consider again two subsequent crests of a GW, where the first is emitted at the event $\hat{r}'_s = (t'_s, r'_s\sin(\theta'_s)\cos(\phi'_s), r'_s\sin(\theta'_s)\sin(\phi'_s), r'_s\cos(\theta'_s))$ and the second at the event $\hat{r}'_{s+\Delta} = (t'_s + \Delta t'_s, r'_s\sin(\theta'_s)\cos(\phi'_s), r'_s\sin(\theta'_s)\sin(\phi'_s), r'_s\cos(\theta'_s))$. The observer detects the first crest at the event $\hat{r}_o = (t_o,0,0,0)$ and the second at the event $\hat{r}_{o+\Delta} = (t_o+\Delta t_o,0,0,0)$. To determine the relativistic redshift we need to solve the integrals we get from setting the metric in \eq{eq:mink} equal to zero. However, this step requires the boundaries of the integrals to be expressed in the same frame and thus we transform $\hat{r}'_s$ and $\hat{r}'_{s+\Delta}$ into the observer frame using the inverse Lorentz transformation, i.e., $\hat{r}_s = \Lambda(-v_z)\hat{r}'_s$ and $\hat{r}_{s+\Delta} = \Lambda(-v_z)\hat{r}'_{s+\Delta}$~\cite{schutz_2009}.

In the observer frame, we get for the emission time of the first crest
\begin{equation}\label{eq:tce1}
    t_s = \gamma(t'_s + vr'_s\cos(\theta'_s))
\end{equation}
and for the emission time of the second crest
\begin{equation}\label{eq:tce2}
    t_{s+\Delta} = \gamma(t'_s + \Delta t'_s + vr'_s\cos(\theta'_s)),
\end{equation}
where $\gamma = (1-v^2)^{-1/2}$ is the Lorentz factor. Furthermore, we have that the radial coordinate of the emitter changes from
\begin{equation}\label{eq:rce1}
    r_s = \sqrt{{r'_s}^2\sin^2(\theta'_s) + \gamma^2(r'_s\cos(\theta'_s) + vt'_s)^2}
\end{equation}
to
\begin{equation}\label{eq:rce2}
    r_{s+\Delta} = \sqrt{{r'_s}^2\sin^2(\theta'_s) + \gamma^2(r'_s\cos(\theta'_s) + v(t'_s + \Delta t'_s))^2},
\end{equation}
while the polar angle changes from
\begin{equation}\label{eq:pce1}
    \cos(\theta_s) = \frac{\gamma(r'_s\cos(\theta'_s) + vt'_s)}{r_s}
\end{equation}
to
\begin{equation}\label{eq:pce2}
    \cos(\theta_{s+\Delta}) = \frac{\gamma(r'_s\cos(\theta'_s) + v(t'_s + \Delta t'_s))}{r_{s+\Delta}},
\end{equation}
and the azimuthal angle remains unaffected. Note that to linear order in $\Delta t'_s$, which is the order to which we expand our results, the difference between $\cos(\theta_s)$ and $\cos(\theta_{s+\Delta})$ scales as $1/r'_s$, and hence it becomes negligible for typical GW sources.

The integrals we need to consider are
\begin{equation}\label{eq:rri1}
    \int_{t_s}^{t_o}\dd t = -\int_0^{r_s}\dd r
\end{equation}
and
\begin{equation}\label{eq:rri2}
    \int_{t_{s+\Delta}}^{t_o + \Delta t_o}\dd t = -\int_0^{r_{s+\Delta}}\dd r.
\end{equation}
Solving the integrals, where we use \eqt{eq:tce1}{eq:rce2} for the boundaries, then subtracting the results of \eq{eq:rri1} from \eq{eq:rri2}, and expanding again to linear order in $\Delta t_o$ and $\Delta t'_s$, we get
\begin{equation}\label{eq:rtd1}
    \Delta t_o = \gamma(1-v\cos(\theta_s))\Delta t'_s,
\end{equation}
where we used \eq{eq:pce1} to simplify the result. Therefore, we recover the classical result for the time dilation due to a relative velocity
\begin{equation}\label{eq:rtd2}
    \dd t_o = \gamma(1-v\cos(\theta))\dd t_s.
\end{equation}
Defining the relativistic redshift as
\begin{equation}\label{eq:drrs}
    (1+z_{\rm rel}) := \gamma(1-v\cos(\theta))
\end{equation}
and using again that the frequency of a GW is the time-derivative of its phase, we get the well-known expression for the relativistic redshift
\begin{equation}\label{eq:rrs}
    f_o = \frac{f_s}{(1+z_{\rm rel})}.
\end{equation}

Next, we derive the apparent distance for a source moving with a relative velocity starting again from \eq{eq:flux}. For a moving source, the area in which the GWs are radiated $\dd A = r^2\dd\Omega$ is affected by a change of the radial coordinate $r$ but also due to a change of the solid angle $\dd\Omega$. For the transformation of the solid angle, we use the transformation of the direction of the radiation described by the wave vector $\hat{k} = (f_o, f_o\sin(\theta)\cos(\phi), f_o\sin(\theta)\sin(\phi), f_o\cos(\theta_o))$. Using that the wave vector transforms as $\hat{k}' = \Lambda(v_z)\hat{k}$, we find
\begin{equation}\label{eq:tpa}
    \cos(\theta') = \frac{\cos(\theta)-v}{1-v\cos(\theta)}.
\end{equation}
From this equation follows
\begin{equation}\label{eq:tdpa}
    \dd\cos(\theta') = \frac{\dd\cos(\theta)}{\gamma^2(1-v\cos(\theta))^2},
\end{equation}
while the azimuthal angle is not affected by the relative velocity ($\dd\phi'=\dd\phi)$. Using $\dd\Omega = \dd\phi\dd\cos(\theta)$ together with \eq{eq:tdpa}, we get
\begin{equation}\label{eq:rsa}
    \dd\Omega = (1+z_{\rm rel})^2\dd\Omega',
\end{equation}
where we used the definition of the relativistic redshift in \eq{eq:drrs}. We find the transformation of the radial coordinate from the vector describing the wavefront of the GW at the time it reaches the observer $\hat{r}_d = (r_d,r_d\sin(\theta)\cos(\phi),r_d\sin(\theta)\sin(\phi),r_d\cos(\theta))$. Using that $\hat{r}'_d = \Lambda(v_z)\hat{r}_d$ and that the radial coordinate equals the magnitude of the spatial vector, we find
\begin{equation}\label{eq:rrc}
    r_d = \frac{r'_d}{(1+z_{\rm rel})}.
\end{equation}

Combining \eqs{eq:rsa}{eq:rrc} with \eq{eq:rtd2} and the fact that the energy of a GW transforms analogous to its frequency (cf. \eq{eq:rrs}) to compute the flux as defined in \eq{eq:flux}, we get
\begin{equation}\label{eq:fadr}
    F = \frac{L}{(1+z_{\rm rel})^2{r'_d}^2\dd\Omega'}.
\end{equation}
Therefore, we find for the apparent distance between the source and the observer
\begin{equation}\label{eq:adrs}
    D_{\rm app} = (1+z_{\rm rel})r'_d.
\end{equation}
It equals the comoving distance at the time the GW reaches the observer times the relativistic redshift, which is in agreement with the result in Ref.~\cite{chen_li_2019}. However, $r'_d$ is the comoving distance in the source frame, and combining \eqs{eq:rrc}{eq:adrs}, we get that the apparent distance in the observer frame equals the comoving distance between times the square of the relativistic redshift
\begin{equation}\label{eq:adro}
    D_{\rm app} = (1+z_{\rm rel})^2r_d.
\end{equation}
Therefore, we find a difference in the effect of a cosmological and a relativistic redshift on the apparent distance. While for an expanding universe the apparent distance scales with a linear factor of the redshift (cf. \eq{eq:adc}), the apparent distance scales with a factor of the relativistic redshift square for a moving source.

\section{Total redshift}\label{sec:trs}

In the two previous, sections we derived the cosmological redshift and its effect on the apparent distance in a flat FLRW universe and the relativistic redshift and how it affects the apparent distance in a Minkowsky universe. In this section, we derive the total redshift for a source moving with a relative velocity in a flat FLRW universe and how the apparent distance changes. We use the FLRW metric in \eq{eq:flrw}, set up COs for the observer $O$ and the source $S$ as in \sref{sec:rrs} but use comoving Lorentz transformation
\begin{equation}\label{eq:clt}
    \Lambda_c(v_z) = \left(\begin{array}{cccc} \gamma & 0 & 0 & -a(t)\gamma v \\
    0 & 1 & 0 & 0 \\ 0 & 0 & 1 & 0 \\ -\gamma v/a(t) & 0 & 0 & \gamma
    \end{array}\right).
\end{equation}
to transform between the two COs to be consistent with the metric. See \aref{app:clt} for a derivation of the comoving Lorentz transformations.

We consider again the case of two subsequent crests of a GW and find for the time coordinate of the first crest
\begin{equation}\label{eq:ctce1}
    t_s = \gamma(t'_s + a(t_s)vr'_s\cos(\theta'_s)),
\end{equation}   
while for the time coordinate of the second crest, we have
\begin{equation}\label{eq:ctce2}
    t_{s+\Delta} = \gamma(t'_s + \Delta t'_s + a(t_{s+\Delta})vr'_s\cos(\theta'_s)),
\end{equation}
where we use the time of the corresponding events in the scale factor. For the radial coordinate of the emitter when the first crest is emitted, we get
\begin{equation}\label{eq:crce1}
    r_s = \sqrt{{r'_s}^2\sin^2(\theta'_s) + \gamma^2(r'_s\cos(\theta'_s) + vt'_s/a(t_s))^2}
\end{equation}
and for the second crest
\begin{align}\label{eq:crce2}
    \nn &r_{s+\Delta} = \\
    &\sqrt{{r'_s}^2\sin^2(\theta'_s) + \gamma^2(r'_s\cos(\theta'_s) + v(t'_s + \Delta t'_s)/a(t_{s+\Delta}))^2}.
\end{align}
The polar angle for the first crest has the form
\begin{equation}\label{eq:cpce1}
    \cos(\theta_s) = \frac{\gamma(r'_s\cos(\theta'_s) + vt'_s/a(t_s))}{r_s}
\end{equation}
while the polar angle for the second crest is
\begin{equation}\label{eq:cpce2}
    \cos(\theta_{s+\Delta}) = \frac{\gamma(r'_s\cos(\theta'_s) + v(t'_s + \Delta t'_s)/a(t_{s+\Delta}))}{r_{s+\Delta}},
\end{equation}
and the azimuthal angle remains unaffected again. We have one more time that to linear order in $\Delta t'_s$, the difference between $\cos(\theta_s)$ and $\cos(\theta_{s+\Delta})$ scales as $1/r'_s$, and hence we can neglect their difference for typical GW sources.

We compute the path of the crests in a FLRW universe by setting $\dd s^2 = 0$ in \eq{eq:flrw}, to obtain
\begin{equation}\label{eq:tri1}
    \int_{t_s}^{t_o}\frac{\dd t}{a(t)} = -\int_0^{r_s}\dd r
\end{equation}
and
\begin{equation}\label{eq:tri2}
    \int_{t_{s+\Delta}}^{t_o+\Delta t_o}\frac{\dd t}{a(t)} = -\int_0^{r_{s+\Delta}}\dd r,
\end{equation}
where the boundaries are given in \eqt{eq:ctce1}{eq:crce2}. Solving the integrals, subtracting the results of \eq{eq:tri1} from \eq{eq:tri2}, and expanding to linear order in $\Delta t_o$ and $\Delta t'_s$, we find
\begin{equation}\label{eq:ttd1}
    \frac{\Delta t_o}{a(t_o)} - \frac{\gamma\Delta t'_s}{a(t_s)} = -\gamma \frac{v}{a(t_s)}\cos(\theta_s)\Delta t'_s,
\end{equation}
where we used \eqs{eq:ctce1}{eq:cpce1} to simplify the expression. Next, we use that $a(t_o)=1$~\cite{dodelson_2020} and that $a(t_s) = (1+z_{\rm cos})^{-1}$, to get
\begin{equation}\label{eq:ttd2}
    \dd t_o = (1+z_{\rm cos})\gamma(1 - v\cos(\theta))\dd t_s.
\end{equation}

Using the definition of the relativistic redshift in \eq{eq:drrs} and that the frequency is the time derivative of the phase, we get
\begin{equation}\label{eq:trs}
    f_o = \frac{f_s}{(1+z_{\rm cos})(1+z_{\rm rel})},
\end{equation}
thus finding that the total redshift is equal to the product of the cosmological redshift and the relativistic redshift
\begin{equation}\label{eq:dtrs}
    (1+z_{\rm tot}) := (1+z_{\rm cos})(1+z_{\rm rel}).
\end{equation}
Our result agrees with Ref.~\cite{bonvin_caprini_2017}, which also considered the combined effect of relative velocity and cosmological expansion on the frequency of GWs, although their approach is restricted to low relative velocities. Therefore, our result represents an extension of their work.

Finally, we derive the combined effect of cosmological expansion and the relative velocity of the source on the apparent distance of a GW source. To get the transformation of the radial coordinate, we consider again the transformation of the wave vector describing the wavefront. To consider the effect of the cosmological expansion, we need to multiply $r_d$ by $a(t_o)$ while we take $r'_d$ times $a(t_s)$. To include the effect of the relative velocity, we use the comoving Lorentz transformation. However, note that because we transform the wave vector seen by the observer (cf. before \eq{eq:rrc}) we have to use $t_o$ in the scale factor in the comoving Lorentz transformation. Therefore, using that $a(t_o)=1$, we get
\begin{equation}\label{eq:crrc}
    r_d = \frac{r'_d}{(1+z_{\rm cos})(1+z_{\rm rel})}.
\end{equation}
where we, further, used that $a(t_s) = (1+z_{\rm cos})^{-1}$. For the solid angle, we use that it is not affected by the cosmological expansion and hence it transforms as in \eq{eq:rsa}.

Inserting these results in \eq{eq:flux} together with the transformation of the time in \eq{eq:ttd2} and using that the energy transforms in the same way as the frequency in \eq{eq:trs}, we get
\begin{equation}\label{eq:fadt}
    F = \frac{L(1+z_{\rm cos})^2}{(1+z_{\rm tot})^2{r'_d}^2\dd\Omega'}
\end{equation}
Therefore, we find for the apparent distance
\begin{equation}\label{eq:adt}
    D_{\rm app} = (1+z_{\rm tot})(1+z_{\rm rel})r_d,
\end{equation}
where we used \eq{eq:crrc} to express the result in the coordinates of the observer. We see that the apparent distance of a moving source in an expanding universe also contains an additional factor of $(1+z_{\rm rel})$. Furthermore, we see that the apparent distance of a source at rest in an expanding universe in \eq{eq:adc} is equal to the one in \eq{eq:adt} for a vanishing relative velocity while the apparent distance of a moving source in a Minkowsky space in \eq{eq:adro} can be obtained by considering a source at a cosmological redshift of zero, where the FLRW metric and the Minkowsky metric become equal.

\section{Observables for high redshift and fast moving sources}\label{sec:obs}

As we showed in the previous sections, the relative velocity of a GW source affects its frequency and apparent distance. In particular, we find that the apparent distance for moving sources needs to be corrected by an additional factor that does not appear in the case of a source at rest relative to the expanding universe at cosmological distances. Therefore, in this section, we analyze how the chirp mass and the apparent distance inferred from observations are affected when the observer ignores the relative velocity of the source.

Before analyzing the effect of the relative velocity, we discuss the observables of GWs which are the amplitude of the wave $h_o$ (in two polarizations), the frequency of the wave $f_o$, and its time derivative $\dot{f}_o$~\cite{sathyaprakash_schutz_2009,chen_li_2019}. From the measurement of $f_o$ and $\dot{f}_o$ in the inspiral part of the signal, where the two merging compact objects are far apart and can be approximated as point masses, one can determine the chirp mass of the source in the observer's frame as
\begin{equation}\label{eq:cmd}
    \mathcal{M}_o := \left(\frac{5\dot{f}_of^{-11/3}_o}{96\pi^{8/3}}\right)^{3/5}.
\end{equation}
It theoretically depends on the masses of the two compact objects as $\mathcal{M}_o = (m_{1,o}m_{2,o})^{3/5}(m_{1,o}+m_{2,o})^{-1/5}$ and describes how the frequency increases with time. Combining this result with the amplitude of the wave $h_o$, one can further determine the apparent distance of the source, using
\begin{equation}\label{eq:ado}
    D_{\rm app} = (\pi f_o\mathcal{M}_o)^{2/3}\frac{4\mathcal{M}_o}{h_o}.
\end{equation}

From Eqs.~(\ref{eq:crs}), (\ref{eq:rrs}), and (\ref{eq:trs}) we see the frequency of a GW transforms $f_o = f_s/(1+z)$ between the observer's and the source's frame, where $(1+z)$ can be either the cosmological, the relativistic, or the total redshift, while from Eqs.~(\ref{eq:ctd2}), (\ref{eq:rtd2}), and (\ref{eq:ttd2}) it follows that $\dot{f}_o = \dot{f}_s/(1+z)^2$. Using these results in \eq{eq:cmd}, we get
\begin{equation}\label{eq:rscm}
    \mathcal{M}_o = (1+z)\mathcal{M}_s,
\end{equation}
where $\mathcal{M}_s := ((5\dot{f}_sf^{-11/3}_s)/(96\pi^{8/3}))^{3/5}$ is the chirp mass in the source's frame. Therefore, we have that the chirp mass of a source redshifted by a factor $(1+z)$ is changed by the same factor. This behavior is known as the `mass-redshift degeneracy' of GWs.

From \eq{eq:ado}, we get that the observed amplitude of a GW goes as
\begin{equation}\label{eq:amp}
    h_o = (\pi f_o\mathcal{M}_o)^{2/3}\frac{4\mathcal{M}_o}{D_{\rm app}}.
\end{equation}
Using that a source at rest and a cosmological redshift ${(1+z_{\rm cos})}$ has an apparent (luminosity) distance $D_{\rm app} = (1+z_{\rm cos})r_d$ (cf. \eq{eq:adc}), the observed chirp mass in \eq{eq:rscm}, and \eq{eq:crs}, we get for the observed amplitude
\begin{equation}\label{eq:oac}
    h_{o,\text{rest}} = (\pi f_s\mathcal{M}_s)^{2/3}\frac{4\mathcal{M}_s}{r_d} =: h_s.
\end{equation}
Therefore, we recover the classical result that the amplitude of GWs is independent of the cosmological redshift~\cite{schutz_1986}.

The amplitude of a GW from a source that is moving with a relative velocity can be determined from \eq{eq:amp} using the apparent distance in the observer frame in \eq{eq:adro} or \eq{eq:adt} for a source in the local universe or at a cosmological distance, respectively. In both cases, we get that the observed amplitude has the form
\begin{equation}\label{eq:oar}
    h_{o,\text{moving}} = \frac{h_s}{(1+z_{\rm rel})}.
\end{equation}
Therefore, we find that the amplitude of a GW from a source with a relative velocity differs from the amplitude of a source at rest by a factor $(1+z_{\rm rel})^{-1}$ independent of its cosmological redshift.

The difference in the amplitude of a GW from a moving source and a source at rest can be understood from the fact that the solid angle is not invariant under (comoving) Lorentz transformations or, to put in a more physical picture, that the aberration of GWs leads to a focusing of the wave rays parallel to the relative velocity of the source~\cite{torres-orjuela_chen_2019,torres-orjuela_chen_2020,torres-orjuela_chen_2021}. The same effect is well known for light as a part of the so-called `headlight effect' or `relativistic beaming' and leads to an additional factor of $(1+z_{\rm rel})^2$ in the intensity of the light~\cite{johnson_teller_1982}.

We have now all tools to analyze the impact of a relative velocity of the source on the chirp mass and the apparent distance inferred from GW detection. Different astrophysical models predict a variety of relative velocities for GW sources, where we consider the following three particularly interesting cases: (i) gravitational kicks which in average have a velocity of $400\,{\rm km\,s^{-1}}$~\cite{amaro-seoane_chen_2016} while extreme kicks of up to $5000\,{\rm km\,s^{-1}}$ can occur for some special configurations~\cite{pretorius_2005,herrmann_hinder_2007a,herrmann_hinder_2007b,herrmann_hinder_2007c,koppitz_pollney_2007,baker_centrella_2006,baker_boggs_2008,campanelli_lousto_2006,campanelli_lousto_2007a,campanelli_lousto_2007b,healy_herrmann_2009}, (ii) BBHs, BNSs, and BHNSBs orbiting a SMBH in the center of a galaxy~\cite{antonini_perets_2012,mckernan_ford_2012,bartos_kocsis_2017,stone_metzger_2017,chen_han_2018,chen_li_2019,addison_gracia-linares_2019,tagawa_haiman_2020,pan_lyu_2021,derdzinski_mayer_2022}, where the velocity depends on the mass of the central SMBH and the radius of the orbit -- for a SMBH with a typical mass of $10^7\,{\rm M_\odot}$ and for an orbital radius of $5\,{\rm mpc}$ the orbital velocity is around $3000\,{\rm km\,s^{-1}}$ while for an orbital radius of $100\,R_s$ ($R_s = 2GM/c^2$ is the Schwarzschild radius of a BH with mass $M$) we have a velocity of around 7\,\% the speed of light, and (iii) the peculiar velocity of galaxies induced by the gravitational interaction of the host galaxy with its environment which in average has been measured to be around $1500\,{\rm km\,s^{-1}}$~\cite{zinn_west_1984,bahcall_1988,girardi_fadda_1996,carlberg_yee_1996,springel_white_2001,ruel_bazin_2014,scrimgeour_davis_2016,colin_mohayaee_2017}. Note that, in principle, these velocities are time-dependent and that the effect of the relative velocity on the redshift also depends on its orientation relative to the line-of-sight (LOS). However, we only want to show the principles of this effect, wherefore we assume the relative velocity to be constant and either parallel or anti-parallel to the LOS.

For a moving source where the observer assumes it to be at rest relative to the expansion of the universe, the error in the chirp mass can be estimated from the relative difference between the chirp mass the source would have if it would be at rest $\mathcal{M}_r = (1+z_{\rm cos})\mathcal{M}_s$ and the chirp mass observed for the moving source $\mathcal{M}_o = (1+z_{\rm tot})\mathcal{M}_s$
\begin{equation}\label{eq:dcm}
    \delta\mathcal{M} := \left|\frac{\mathcal{M}_r - \mathcal{M}_o}{\mathcal{M}_o}\right| = \left|1 - \sqrt{\frac{1 \mp v}{1 \pm v}}\right|,
\end{equation}
where the upper sign corresponds to the relative velocity being anti-parallel to the LOS (relativistic redshift) and the lower sign to the relative velocity being parallel to the LOS (relativistic blueshift).

\begin{figure}[tpb] \centering \includegraphics[width=0.48\textwidth]{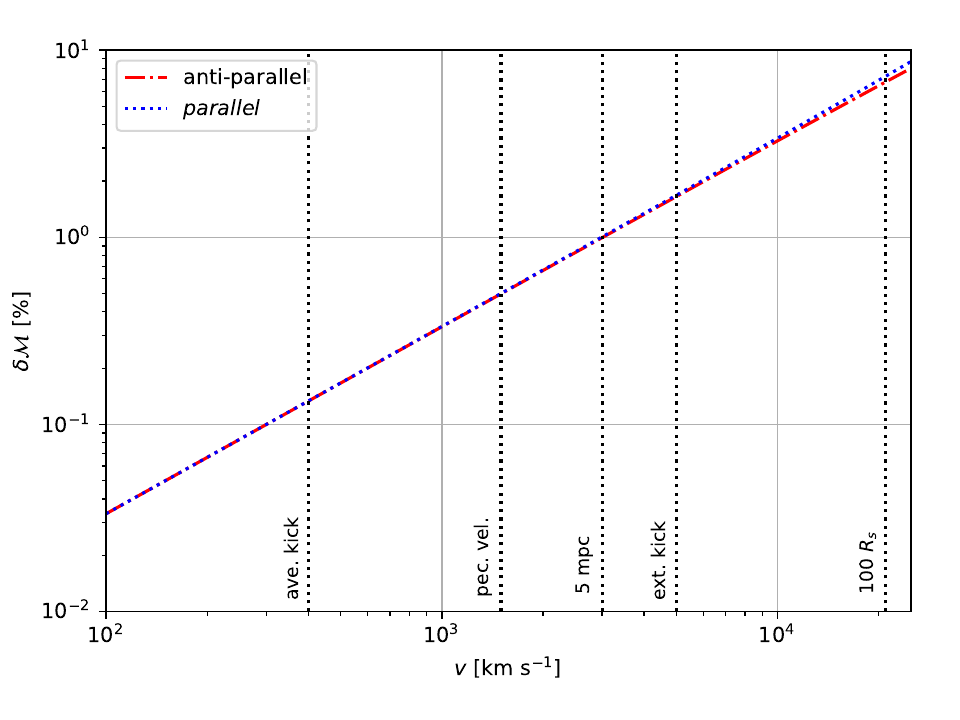}
\caption{
    The relative error in the chirp mass for a source moving at different relative velocities when the observer assumes the source is at rest. The red dashed-dotted line corresponds to the relative velocity and the LOS being anti-parallel while the blue dotted line corresponds to the two being parallel. The vertical lines represent typical velocities for the astrophysical scenarios described in the text.
    }
\label{fig:ecm}
\end{figure}

\fref{fig:ecm} shows the relative error in the measurement of the chirp mass of the source when the observer assumes it is at rest for different relative velocities. We see that the error is a monotonic function of the relative velocity and that it can reach 1\,\% for a velocity of $3000\,{\rm km\,s^{-1}}$ corresponding to a binary orbiting a SMBH of the mass $10^7\,{\rm M_\odot}$ at a distance of $5\,{\rm mpc}$ while for lower velocities like the average kick velocity and the peculiar velocity of galaxies the error is below 1\,\%. For extreme kicks of $5000\,{\rm km\,s^{-1}}$ the error is slightly higher than 1.5\,\% and it can go up to around 7\,\% for a binary orbiting a SMBH at $100\,R_s$. We further can see that there is a difference in the error when the relative velocity and the LOS are anti-parallel (source moving away from the observer) or parallel (source approaching the observer) but the difference is relatively small and only becomes relevant for highly relativistic velocities of several percent the speed of light.

The error in the apparent distance can be estimated considering the relative difference between the apparent distance $D_{{\rm app},r} = (1+z_{\rm cos})D_s$ ($D_s := (\pi f_s\mathcal{M}_s)^{2/3}(4\mathcal{M}_s/h_s))$ the source would have if it would be at rest and the observed apparent distance for the moving source $D_{{\rm app},o} = (1+z_{\rm tot})(1+z_{\rm rel})D_s$
\begin{equation}\label{eq:ead}
    \delta D_{\rm app} := \left|\frac{D_{{\rm app},r} - D_{{\rm app},o}}{D_{{\rm app},o}}\right| = \left|\frac{2v}{1 \pm v}\right|,
\end{equation}
where the plus sign corresponds to the relative velocity being anti-parallel to the LOS and the minus sign to the relative velocity and the LOS being parallel.

\begin{figure}[tpb] \centering \includegraphics[width=0.48\textwidth]{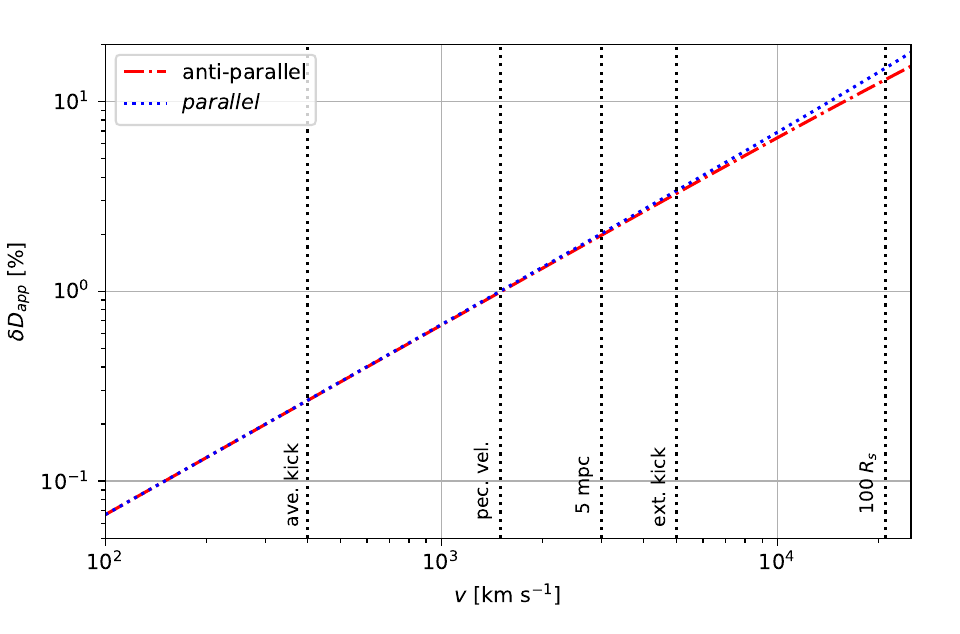}
\caption{
    The relative error in the apparent distance for a source moving at different relative velocities when the observer assumes the source is at rest. The lines represent the same cases as in \fref{fig:ecm}.
    }
\label{fig:ead}
\end{figure}

In \fref{fig:ead}, we see that the relative error in the measurement of the apparent distance when the observer assumes the source is at rest is bigger than for the chirp mass. The error is still negligible for average kick velocities of $400\,{\rm km\,s^{-1}}$ but for a peculiar velocity of galaxies of $1500\,{\rm km\,s^{-1}}$ the error can go up to around 1\,\%. For a binary orbiting a SMBH of mass $10^7\,{\rm M_\odot}$ at a distance of $5\,{\rm mpc}$ the error is above 2\,\% and for extreme kicks the error goes up to around 3\,\%. For highly relativistic velocities the dependence on the orientation of the velocity relative to the LOS becomes again more prominent so that for a binary orbiting a SMBH at a radius of $100\,R_s$ and the relative velocity being anti-parallel to the LOS the error in the apparent distances is around 13\,\% while for the relative velocity and the LOS being parallel the error increases to 15\,\%.

We see that ignoring the relative velocity of the source impacts the measurement of the chirp mass and the apparent distance causing errors of up to 7\,\% in the first and of up to 15\,\% in the latter. Therefore, including a relative velocity of the source is crucial to obtain accurate results and not over- or underestimating its parameters, where the apparent distance is particularly affected due to its $(1+z_{\rm rel})^2$-dependence. Moreover, the apparent distance plays a crucial role in determining the Hubble constant $H_0$ using standard sirens~\cite{schutz_1986}, wherefore we analyze the impact of the relative velocity in measurements of $H_0$ in the following section.

\section{The effect of the source's relative velocity on the measurement of the Hubble constant}\label{sec:hub}

GW sources with EM counterparts have been proposed as standard sirens to measure the Hubble constant $H_0$~\cite{schutz_1986,holz_hughes_2005}. Up to date, the two most prominent GW events used for measurements of $H_0$ are GW170817, the merger of a BNS with a subsequent optical transient~\cite{ligo_2017c,ligo_2017d,ligo_2017e,ligo_2017f,chen_fishbach_2018}, and GW190521, a merger of a BBH with the event ``ZTF19abanrhr'' as a candidate EM counterpart~\cite{mukherjee_ghosh_2020,chen_haster_2022}. Usually, the sources are assumed to be at rest relative to the expanding universe although there are many astrophysical scenarios where the source needs to be considered as moving. In Refs.~\cite{wang_wang_2018,howlett_davis_2020,nicolaou_lahav_2020,mukherjee_lavaux_2021} the effect of a source's relative velocity, in particular the peculiar velocity of the host galaxy, has been included to correct for the error on the redshift of the EM counterpart, however, the effect of the relative velocity on the amplitude respective the square factor for $(1+z_{\rm rel})$ in the apparent distance has been ignored. Therefore, two different scenarios for errors need to be considered when analyzing the impact of the source's relative velocity on the measurement of the Hubble constant: (i) the effect of the relative velocity is considered in the redshift of the source but not for its apparent distance, and (ii) the source is moving but the relative velocity is completely ignored. Case (i) is a common error in most studies including the effect of a relative velocity. Case (ii) arises when a wrong astrophysical model for the source's relative velocity is being applied, e.g., when using the usual assumption that the source only moves due to the peculiar velocity of its host galaxy but in reality, it has a much higher velocity induced by an orbital motion around a SMBH as discussed in Ref.~\cite{peng_chen_2021}, where it is shown that 1-2\,\% of all LIGO/Virgo/KAGRA sources merge at a distance of $10\,R_s$ from a SMBH.

For a source at a small cosmological distance ($z_{\rm cos} \ll 1$), we have the following relation between its cosmological redshift $z_{\rm cos}$, its luminosity distance $D_L$, and the Hubble constant $H_0$~\cite{maggiore_2008}
\begin{equation}\label{eq:rlh}
    z_{\rm cos} = H_0D_L.
\end{equation}
Using that $D_L := (1+z_{\rm cos})r_d$, where $r_d$ is the comoving distance between source and observer in the frame of the observer, we further get $z_{\rm cos} = H_0r_d/(1-H_0r_d)$. From \eq{eq:rlh} it follows that $H_0$ can be inferred using the apparent distance of the GW source $D_{\rm app}$ and the redshift $z_{\rm EM}$ of its EM counterpart as
\begin{equation}\label{eq:hub}
    H_{0,i} = \frac{z_{\rm EM}}{D_{\rm app}}.
\end{equation}
However, if the source is moving the effect of the relative velocity on the redshift and the apparent distance needs to be corrected, otherwise we would infer a wrong value for $H_0$. As mentioned above, a typical cause for the relative velocity of the source is the peculiar velocity of its host galaxy~\cite{wang_wang_2018,howlett_davis_2020,nicolaou_lahav_2020,mukherjee_lavaux_2021}. For BNS mergers the EM counterpart comes from a subsequent kilonova produced by the interaction of ejected mass interacting with matter surrounding the newly formed compact object~\cite{metzger_2017,arcavi_hosseinzadeh_2017,smartt_chen_2017,villar_guillochon_2017}. For BBHs the EM counterpart can be produced through different mechanisms like the interaction with the surrounding gas in the center of a galaxy due to the gravitational kick of the remnant black hole or a jet ejected during the merger, and BBHs merging inside a common envelope followed by an explosive EM counterpart~\cite{graham_ford_2020,perna_lazzati_2021,ginat_glanz_2020}. In general, the relative velocity of the GW source and its EM counterpart can differ but, in most cases, the velocity of the GW source and the EM counterpart are equal or at least of the same order; thus we consider them to be equal in our analysis.

\begin{figure}[tpb] \centering \includegraphics[width=0.48\textwidth]{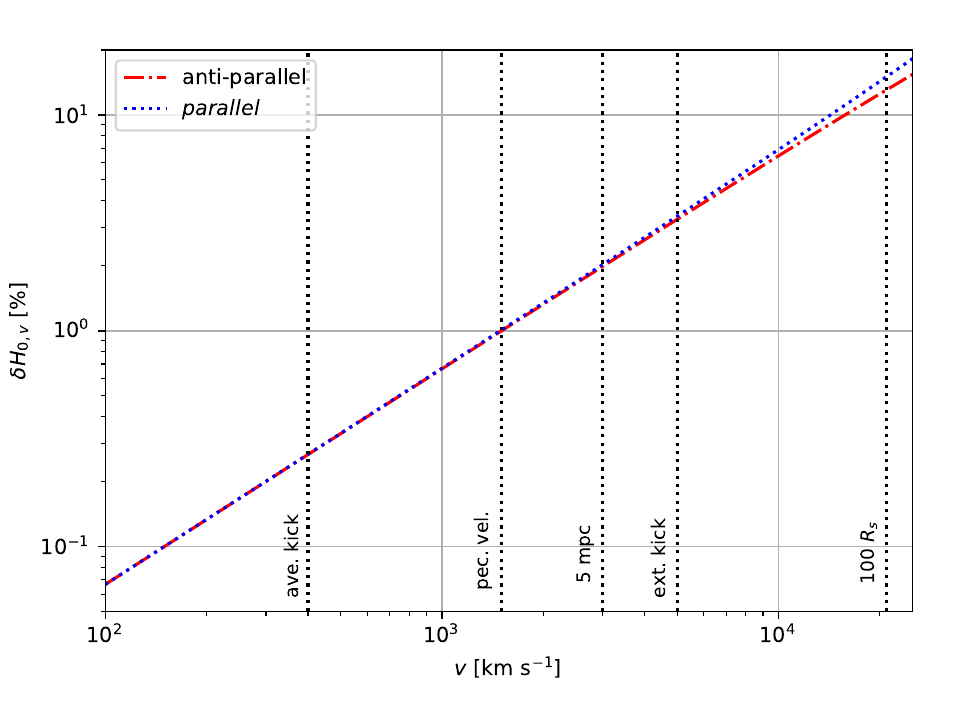}
\caption{
    The relative error in the Hubble constant for a source moving at different relative velocities when the observer only considers the effect of the relative velocity on the redshift of the EM counterpart but not the apparent distance of the source respective the amplitude of the GW. The lines correspond to the same cases as in \fref{fig:ecm}.
    }
\label{fig:ehcv}
\end{figure}

An observer including the effect of the relative velocity on the redshift but ignoring its effect on the apparent distance would use $z_{\rm EM} = z_{\rm cos}$ to infer the Hubble constant, thus finding
\begin{equation}\label{eq:hiv}
    H_{0,v} := \frac{z_{\rm tot}}{D_{\rm app}} = \frac{H_0}{(1+z_{\rm rel})^2},
\end{equation}
where $(1+z_{\rm rel})$ is in \eq{eq:drrs}. Therefore, we recover the correct value for $H_0$ if the source is at rest but if the source is moving with a relative velocity we obtain an erroneous result. The error done by not considering the effect of the relative velocity on the apparent distance can be estimated as
\begin{equation}\label{eq:ehiv}
    \delta H_{0,v} := \left|\frac{H_{0,v} - H_0}{H_0}\right| = \left|\frac{2v}{1 \pm v}\right|,
\end{equation}
where we again consider the special cases of the relative velocity being anti-parallel to the LOS (plus sign) and parallel to the LOS (minus sign). We see in \fref{fig:ehcv}, that the error increases with an increasing magnitude of the relative velocity and already can reach around 1\,\% for a peculiar velocity of the host galaxy of $1500\,{\rm km\,s^{-1}}$. For a binary orbiting a SMBH of $10^7\,{\rm M_\odot}$ at a distance of $5\,{\rm mpc}$ with a velocity of $3000\,{\rm km\,s^{-1}}$ the error goes up to around 2\,\% and can exceed 3\,\% for extreme kicks of $5000\,{\rm km\,s^{-1}}$. For higher relative velocities the orientation of the velocity relative to the LOS becomes significant, varying from 13\,\% for the two being anti-parallel to 15\,\% for the two being parallel for a binary orbiting a SMBH at a distance of $100\,R_s$.

An observer completely ignoring the effect of relative velocity would infer the Hubble constant using the total redshift for the EM counterpart ($z_{\rm EM} = z_{\rm tot}$) and the apparent luminosity distance, thus obtaining
\begin{equation}\label{eq:hir}
    H_{0,r} := \frac{z_{\rm tot}}{D_{\rm app}} = \frac{(1+z_{\rm cos})z_{\rm rel} + H_0D_L}{(1+z_{\rm rel})^2D_L}.
\end{equation}
We recover again the correct value for $H_0$ if the source is at rest but, if the source is moving with a relative velocity, we obtain an erroneous result and, in particular, the value inferred for $H_0$ depends on the distance of the source. The error done by ignoring the relative velocity of the source can be estimated by
\begin{align}\label{eq:ehir}
    \nn \delta H_{0,r} :=& \left|\frac{H_{0,r} - H_0}{H_0}\right| \\
    =& \left|\frac{(1+z_{\rm rel})^2z_{\rm cos} - (1+z_{\rm rel})(1+z_{\rm cos}) + 1}{(1+z_{\rm rel})^2z_{\rm cos}}\right|,
\end{align}
where we used \eq{eq:rlh} to eliminate the direct dependence on $H_0$. We, furthermore, use that $(1+z_{\rm rel}) = \sqrt{(1\pm v)/(1\mp v)}$ for the relative velocity and the LOS being either anti-parallel (upper sign) or parallel (lower sign).

\begin{figure}[tpb] \centering \includegraphics[width=0.48\textwidth]{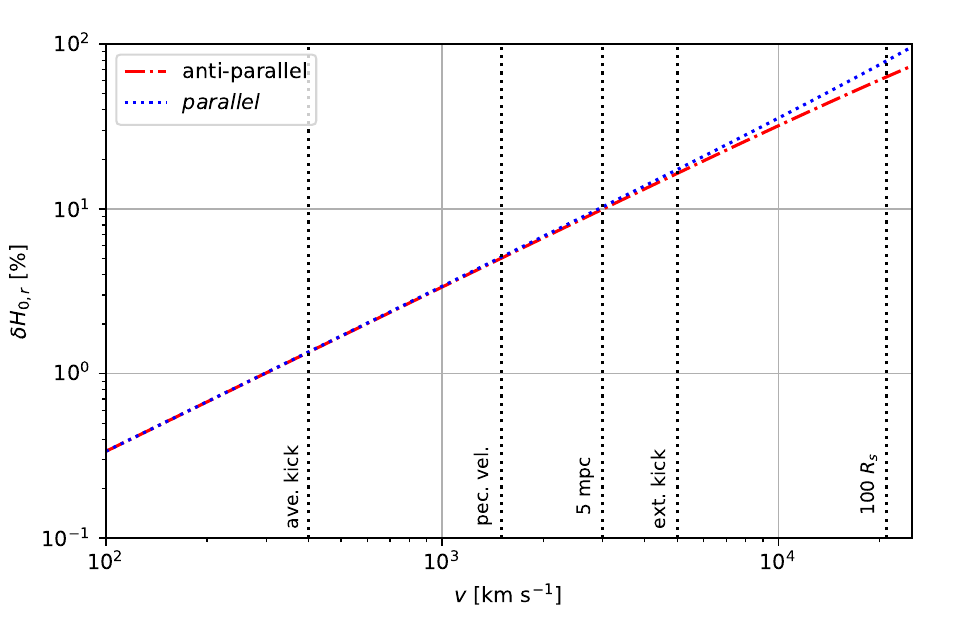}
\caption{
    The relative error in the Hubble constant for a source at the cosmological distance of GW150914 ($z_{\rm cos}=0.09$) that is moving at different relative velocities but the observer assumes it to be at rest relative to the expanding universe. The lines correspond to the same cases as in \fref{fig:ecm}.
    }
\label{fig:ehcrv}
\end{figure}

\begin{figure}[tpb] \centering \includegraphics[width=0.48\textwidth]{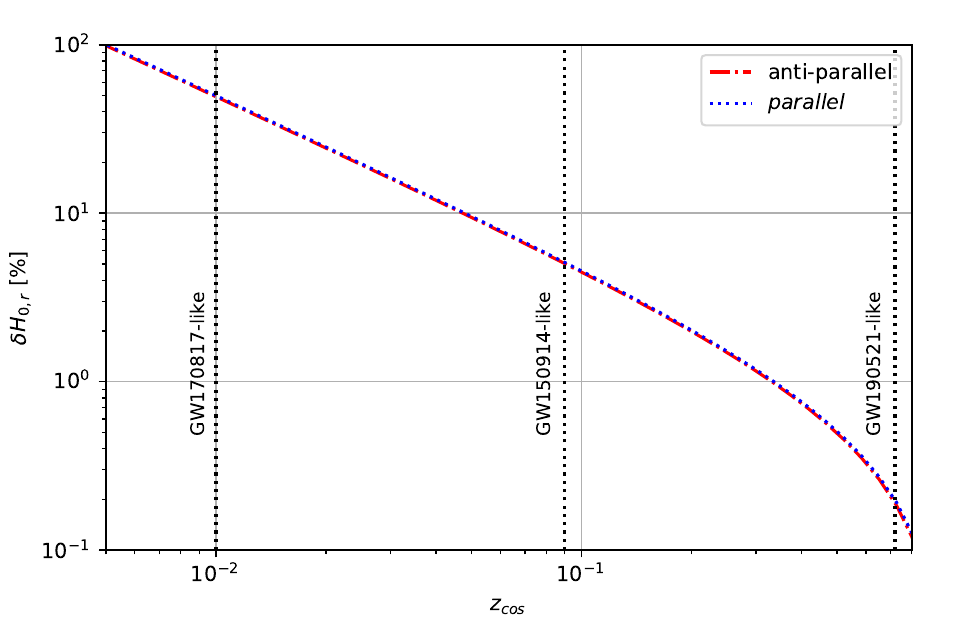}
\caption{
    The relative error in the Hubble constant for a source moving with a typical peculiar velocity of galaxies of $1500\,{\rm km\,s^{-1}}$ for different cosmological redshifts of the source. The red dashed-dotted line corresponds to the relative velocity and the LOS being anti-parallel while the blue dotted line corresponds to the two being parallel. The vertical lines correspond to the cosmological redshift for the GW events GW170817, GW150914, and GW190521.
    }
\label{fig:ehcrc}
\end{figure}

The error in the measurement of the Hubble constant further increases when the effect of the relative velocity is not only ignored in the apparent distance but also the redshift of the EM counterpart. \fref{fig:ehcrv} shows the relative error for a source at a cosmological redshift of $0.09$ (corresponding to the distance of GW150914~\cite{GWTC1}) for different velocities. We see that even for relatively low velocities of only $400\,{\rm km\,s^{-1}}$, corresponding to the average kick velocity, the error can reach around 1.5\,\%. For an average peculiar velocity of the host galaxy of $1500\,{\rm km\,s^{-1}}$ the error already exceeds 5\,\%, and the error becomes bigger than 10\,\% for a source moving with more than $3000\,{\rm km\,s^{-1}}$, corresponding to the orbital velocity of a binary orbiting a SMBH of $10^7\,{\rm M_\odot}$ at $5\,{\rm mpc}$. The dependence on the orientation of the velocity relative to the LOS becomes again more prominent for high relative velocities thus causing the error to vary between around 60\,\% and 80\,\% for a binary at an orbit of $100\,R_s$ around a SMBH. In \fref{fig:ehcrc}, we fix the relative velocity of the source to be the typical peculiar velocity of galaxies and vary the cosmological redshift of the source. We see that for this relatively low velocity the error is almost the same for the relative velocity being either anti-parallel or parallel to the LOS while it decreases for an increasing cosmological redshift. For a source at the distance of GW170817 ($z_{\rm cos}=0.01$)~\cite{GWTC1}, the error in $H_0$ is more than 50\,\%. The error for a source at the distance of GW150914 decreases to still considerable 5\,\% and goes below 1\,\% for sources at cosmological redshifts of more than 0.3. We see that for a source at the distance of GW190521 ($z_{\rm cos}=0.72$)~\cite{GWTC2} the error goes down to around 0.2\,\%. However, it should be noted that \eq{eq:rlh} assumes the cosmological redshift to be much smaller than one and hence the estimation for redshifts close to one are less reliable. In general, we expect the error for high cosmological redshifts to be at least bounded from below by the errors found when only correcting for the effect of the relative velocity on the redshift of the EM counterpart, which is independent of the cosmological redshift.

Last, we consider the case of GW170817 using a relative velocity of roughly $300\,{\rm km\,s^{-1}}$ as used in the analysis conducted by the LIGO and Virgo Collaborations~\cite{ligo_2017c,ligo_2017d}. From \eq{eq:ehir} we get that if the relative velocity of the source would be ignored completely the error would be around 100\,\% thus making the result useless. When correcting by the effect of the relative velocity on the redshift of the EM counterpart but ignoring its effect on the apparent distance as done in Ref.~\cite{ligo_2017d}, the error can be estimated using \eq{eq:ehiv} to be 0.2\,\% and hence we confirm that their result is reliable while the error in $H_0$ is dominated by other effects. However, for sources with higher relative velocities the error when ignoring their effect on the apparent distance increases quickly, hence it needs to be considered.

Although the orientation of the relative velocity is random the error induced by not properly considering the relative velocity is not symmetric in the orientation and hence we do not expect the error to cancel out for a high number of detections. Moreover, the relativistic shift moves heavy (and hence usually loud) BBHs that are redshifted towards the more sensitive frequency range of LIGO/Virgo/KAGRA while blueshifted sources are moved out of band. This shift should induce a selection effect that will additionally hinder a reduction of the error by an increase in numbers. Similar selection effects may also occur for other sources and other detectors.

\section{Conclusions}\label{sec:con}

We studied the effect of cosmological expansion and the relative velocity of the source on the frequency of GWs and the apparent distance obtained from them. We recover the classical result that the frequency of a source at rest relative to the expanding universe is redshifted by the cosmological expansion while its apparent distance increases by the same redshift factor. We also show that the total redshift of a source moving with a relative velocity in an expanding universe is equal to the product of the cosmological redshift times the relativistic redshift, as usually assumed. However, we find that the apparent distance of a source moving with a relative velocity changes by a factor of the relativistic redshift square, in contrast to the linear behavior for the cosmological redshift. The additional factor can be attributed to the fact that the amplitude of GWs from a moving source changes by the same factor relative to a source at rest.

Using these results we analyze the estimation error in the two main observables of GWs, the chirp mass of the source and its apparent distance when the relative velocity is ignored. We consider the three astrophysical scenarios of gravitational kicks, the orbital motion of BBHs, BNSs, or BHNSBs around a SMBH, and the peculiar velocity of the host galaxy which lead to velocities varying from a few $100\,{\rm km\,s^{-1}}$ up to several percent of the speed of light. The error in the chirp mass and the apparent distance varies between 0.1\,\% respective 0.25\,\% for average kick velocities and go up to 7\,\% and 15\,\%, respectively, for sources close to a SMBH. Furthermore, the error exceeds 1\,\% in the apparent distance for a source moving due to the peculiar velocity of its host galaxy. Therefore, the errors induced by a relative velocity are at a similar level to current estimation errors for sources close to a SMBH but there is no critical effect for lower velocities~\cite{GWTC1,GWTC2,GWTC3}.

Furthermore, we study how the relative velocity of the source impacts the estimation of the Hubble constant using standard sirens when only considering the effect of the relative velocity on the redshift of the EM counterpart but not on the apparent distance and when completely ignoring the effect of the source's relative velocity. We find that in the first case, the error in $H_0$ can already reach 1\,\% for a typical peculiar velocity of the host galaxy and go up to 15\,\% for a binary close to a SMBH. When completely ignoring the effect of the relative velocity, the error increases to 5\,\% for a typical peculiar velocity of the host galaxy for a source at the cosmological redshift of GW150912. We, furthermore, find that the error depends on the cosmological redshift of the source where for a typical peculiar velocity of the host galaxy it can go up to over 50\,\% for a source like GW170817 at a cosmological distance of 0.01 while it decreases to under 1\,\% for sources at cosmological redshifts of more than 0.3.

The measurement of the Hubble constant using standard sirens is affected by estimation errors in the sky localization of the GW source and its EM counterpart, where the error is dominated by instrumental uncertainties at a 10\,\%-level~\cite{holz_hughes_2005,ligo_2017d,GWTC3}. Other errors from the `inclination-distance-degeneracy' or possible lensing of the source add to the systematic error to at least a percent level. However, these errors can be mitigated by proper modeling of the respective effect or, e.g., by breaking the `inclination-distance-degeneracy' when including higher spherical modes~\cite{cantiello_jensen_2018,usman_mills_2019,kumar_blackman_2019}. The error induced by a relative velocity of the source when only considering its effect on the redshift of the EM counterpart is at a similar level to other estimation errors and can be even bigger than these when completely ignored. Therefore, we argue that the relative velocity of the source needs to be modeled as a systematic error, including its effect on the apparent distance of the source, to obtain accurate estimations of the Hubble constant.

\section*{Acknowledgments}

ATO acknowledges support from the Guangdong Major Project of Basic and Applied Basic Research (Grant No. 2019B030302001) and the China Postdoctoral Science Foundation (Grant No. 2022M723676). XC is supported by the National Science Foundation of China (Grants Nos. 11873022 and 11991053).

\appendix
\section{Comoving Lorentz transformations}\label{app:clt}

When considering a flat universe, one expects that the laws of special relativity are valid since there is no gravitational force. This implies that Lorentz transformations can be used to transform between moving observers. However, Lorentz transformations do not preserve the line element for a flat FLRW metric and hence they need to be adapted. In this section, we derive \textit{comoving Lorentz transformations} that are analogous to the usual Lorentz transformations but leave the line element for the flat FLRW metric invariant.

We can write the flat FLRW metric (cf. \eq{eq:flrw}) as
\begin{equation}\label{eq:fflrw}
    \dd s^2 = \dd t^2 - a^2(t)\dd\bs{r}^2,
\end{equation}
while we denote the geodesic of a particle in this space as $\hat{p} = (p_0,\bs{p})$. We see that for such a geodesic the line element has the form $\dd s^2 = \dd p_0^2 - a^2(t)\dd\bs{p}^2$. If we use the regular Lorentz transformation for a velocity along the $z$-coordinate to transform the geodesic $\hat{p}' = \Lambda(v_z)\hat{p}$, we get for the line element
\begin{align}\label{eq:lelt}
    \nn \dd s^2 =& (\gamma^2 - a^2(t)\gamma^2v^2)\dd p_0^2 - 2(1 - a^2(t))\gamma^2v\dd p_0\dd p_z \\
    &- a^2(t)\dd p_x^2 - a^2(t)\dd p_y^2 - (a^2(t)\gamma^2 - \gamma^2v^2)\dd p_z^2
\end{align}
and thus we see the that line element is not invariant under Lorentz transformations.

To find the transformations that leave the line element invariant, we define the following transformation
\begin{equation}\label{eq:dclt}
    \Lambda_c(v_z) = \left(\begin{array}{cccc} A\gamma & 0 & 0 & -B\gamma v \\
    0 & C & 0 & 0 \\ 0 & 0 & D & 0 \\ -E\gamma v & 0 & 0 & F\gamma
    \end{array}\right).
\end{equation}
Note that this is just the Lorentz transformation with additional factors in all non-vanishing elements. Our goal is to determine the factors $A, B, C, D, E, F$ so that the line element remains invariant under these transformations. Transforming the geodesic and inserting the result in \eq{eq:fflrw}, we get
\begin{align}\label{eq:leclt}
    \nn \dd s^2 =& (A^2\gamma^2 - E^2a^2(t)\gamma^2v^2)\dd p_0^2 - 2(AB - EFa^2(t)) \\
    \nn &\times \gamma^2v\dd p_0\dd p_z - C^2a^2(t)\dd p_x^2 - D^2a^2(t)\dd p_y^2 \\
    &- (F^2a^2(t)\gamma^2 - B^2\gamma^2v^2)\dd p_z^2.
\end{align}
Therefore, for the line element to be invariant requires
\begin{subequations}
\begin{align}\label{eq:coe1}
    1 =& A^2\gamma^2 - E^2a^2(t)\gamma^2 + E^2a^2(t), \\ \label{eq:coe2}
    0 =& AB - EFa^2(t), \\ \label{eq:coe3}
    1 =& C^2, \\ \label{eq:coe4}
    1 =& D^2, \\ \label{eq:coe5}
    a^2(t) =& F^2a^2(t)\gamma^2 - B^2\gamma^2 + B^2
\end{align}
\end{subequations}
where we used $\gamma^2v^2 = \gamma^2-1$ to reshape the equations.

From \eqs{eq:coe3}{eq:coe4} and imposing that the transformation shall not invert directions, we get $C=D=1$. We, further, see that for \eq{eq:coe1} to be fulfilled, we need $A=1$ and $E=1/a(t)$. Using this result in \eq{eq:coe2}, we get $B=a(t)F$ which together with \eq{eq:coe5} and the requirement that the transformation does not invert the orientation of the geodesic leads to $F=1$. Therefore, we get for the comoving Lorentz transformation for a velocity in direction of the $z$-coordinate
\begin{equation}\label{eq:clta}
    \Lambda_c(v_z) = \left(\begin{array}{cccc} \gamma & 0 & 0 & a(t)\gamma v \\
    0 & 1 & 0 & 0 \\ 0 & 0 & 1 & 0 \\ \gamma v/a(t) & 0 & 0 & \gamma
    \end{array}\right).
\end{equation}

We highlight that the comoving Lorentz transformation resembles several properties of the regular Lorentz transformation, in particular, that the inverse comoving Lorentz transformation can be obtained by inverting the sign of the velocity. Moreover, the inverse comoving Lorentz transformation also leaves the line element invariant. Last, we point out that the time in the scale factor needs to be the time of the event transformed using the time in the rest frame of the observer.

\bibliographystyle{apsrev4-2}

\end{document}